\documentclass[useAMS,usenatbib]{mn2e}

\usepackage{graphicx}
\usepackage{amssymb}
\usepackage{amsmath}
\usepackage{draftcopy}
\usepackage[normalem]{ulem}
\usepackage{color}

\begin{document}

\title[3D MHD model of H$\alpha$ emission]
{
First 3D MHD simulation of a massive-star magnetosphere
with application to H$\alpha$ emission from \t1oc
}
 \author[A. ud-Doula et al.]
 {A. ud-Doula$^1$\thanks{Email: uddoula@psu.edu}, J. O. Sundqvist $^2$,
 S. P. Owocki $^2$, V. Petit $^3$ and R.H.D. Townsend$^4$ \\
 $^1$  Penn State Worthington Scranton,
 Dunmore, PA 18512, USA.\\
 $^2$ Bartol Research Institute,
     University of Delaware,
     Newark, DE 19716, USA.\\
  $^3$ Department of Geology and Astronomy, West Chester University, West Chester, PA 19383, USA\\   
 $^4$ Department of Astronomy, University of Wisconsin-Madison,
 5534 Sterling Hall, 475 N Charter Street, Madison, WI 53706, USA.
}


\def\<<{{\ll}}
\def\>>{{\gg}}
\def\wig{{\sim}}
\def\spose#1{\hbox to 0pt{#1\hss}}
\def\ltwig{\mathrel{\spose{\lower 3pt\hbox{$\mathchar"218$}}
     \raise 2.0pt\hbox{$\mathchar"13C$}}}
\def\gtwig{\mathrel{\spose{\lower 3pt\hbox{$\mathchar"218$}}
     \raise 2.0pt\hbox{$\mathchar"13E$}}}
\def\+/-{{\pm}}
\def\=={{\equiv}}
\def\Rstar{R_{\ast}}
\def\Mstar{M_{\ast}}
\def\Lstar{L_{\ast}}
\def\Tstar{T_{\ast}}
\def\gstar{g_{\ast}}
\def\vth{v_{th}}
\def\grad{g_{rad}}
\def\glines{g_{\rm{lines}}}
\def\Mdot{\dot M}
\def\mdot{\dot m}
\def\yr{{\rm yr}}
\def\ksec{{\rm ksec}}
\def\kms{{\rm km \, s^{-1}}}
\def\qad{\dot q_{ad}}
\def\qlines{\dot q_{lines}}
\def\solar{\odot}
\def\Msun{M_{\solar}}
\def\msbyr{\Msun/\yr}
\def\Rsun{R_{\solar}}
\def\Lsun{L_{\solar}}
\def\Be{{\rm Be}}
\def\Rpole{R_{p}}
\def\Req{R_{eq}}
\def\Rmin{R_{min}}
\def\Rmax{R_{max}}
\def\Rstag{R_{stag}}
\def\vinf{V_\infty}
\def\Vrot{V_{\rm{rot}}}
\def\t1oc{$\theta^1$~Ori~C}
\def\Vcrit{V_{crit}}
\def\half{{1 \over 2}}
\newcommand{\beq}{\begin{equation}}
\newcommand{\eeq}{\end{equation}}
\newcommand{\beqa}{\begin{eqnarray}}
\newcommand{\eeqa}{\end{eqnarray}}
\def\phip{{\phi'}}

\maketitle

\begin{abstract}

We present the first fully 3D MHD simulation for magnetic channeling and confinement of a radiatively driven, massive-star wind.
The specific parameters are chosen to represent the prototypical slowly rotating magnetic O star \t1oc,
for which centrifugal and other dynamical effects of rotation are negligible.
The computed global structure in latitude and radius resembles that found in previous 2D simulations, with unimpeded outflow along open field lines near the magnetic poles, and a complex equatorial belt of inner wind trapping by closed loops near the stellar surface, giving way  to outflow above the Alfv\'{e}n radius.
In contrast to this previous 2D work, the 3D simulation described here now also shows how this complex structure fragments in azimuth, forming distinct clumps of closed loop infall within the Alfv\'{e}n radius, transitioning in the outer wind to radial spokes of enhanced density with characteristic azimuthal separation of $15-20 \degr$.
Applying these results in a 3D code for  line radiative transfer, we show that emission from the associated  3D `dynamical magnetosphere'  matches well the observed   H$\alpha$ emission seen from \t1oc, fitting both its dynamic spectrum over rotational phase, as well as the observed  level of cycle to cycle stochastic variation.
Comparison with previously developed 2D models for Balmer emission from a dynamical magnetosphere generally confirms that  time-averaging over 2D snapshots can be a good proxy for the spatial averaging over 3D azimuthal wind structure.
Nevertheless, fully 3D simulations will still be needed to model the emission from magnetospheres with non-dipole field components, such as suggested by asymmetric features seen in the H$\alpha$ equivalent-width curve of \t1oc.
\end{abstract}


\section{Introduction}

Massive, hot (OB type) stars lack the vigorous subsurface convection thought to drive the dynamo central to the magnetic activity cycles of the sun and other cool stars. 
Nonetheless, new generations of spectropolarimeters have directly revealed large-scale, organized (often significantly dipolar) magnetic fields ranging in strength from 0.1 to 10 kG in several dozen OB stars 
\citep[e.g.][]{Don2002,Don2006,Hub2006,Pet2008,Gru2009,Mar2010} 
with ongoing monitoring and surveys carried out by the  MiMeS (Magnetism in Massive Stars) consortium  \citep{Wad2012}.

A key theoretical issue, examined in a series of papers \citep{udDOwo2002,udD2008,udD2009}, is how such strong fields can, in combination with stellar rotation, channel and confine the radiatively driven stellar wind outflows of such OB stars, leading to the formation of a circumstellar {\em magnetosphere}.
In relatively rapidly rotating  magnetic Bp stars (with periods of order a day), magnetically trapped wind material above the Kepler co-rotation radius is centrifugally supported, and so builds up into a stable \textit{centrifugal magnetosphere} \citep{TowOwo2005}, often leading to rotationally modulated, high velocity Balmer line emission \citep{Tow2005}.

However, likely because of the angular momentum loss in their stronger magnetic winds,
most O-stars with detected surface magnetic fields are
\textit{slow rotators}, with periods ranging from 15~days for \t1oc to  538~days for HD\,191612 and 55~years for HD~108 \citep{Sta2008, How2007, Mar2010}, 
implying a very limited dynamical effect of rotation on the wind magnetic channeling\footnote{One exception is
Plaskett's star \citep{Gru2012b}, in which mass accretion from the close binary companion may have spun up the inferred magnetic star.}. 
Indeed, \citet{Gag2005} showed that
non-rotating, two-dimensional (2D) magnetohydrodynamic (MHD) simulations match well the temperature,
luminosity, and rotational phase-dependent occultation of the X-ray emitting plasma around \t1oc.

For the even more slowly rotating case of HD\,191612, \citet{Sun2012}  have used similar 2D MHD simulations to show that both the strength and rotational phase variation of the observed H$\alpha$ emission can be well reproduced in terms of a {\em dynamical magnetosphere}; in this case,  the higher O-star mass loss rate means that, even without significant centrifugal support, the transient magnetic suspension and subsequent gravitational infall of magnetically trapped material yields a high enough circumstellar density to give strong Balmer line emission (see also \citealt{Gru2012} for a similar application to another O-star  HD\,57682).

To provide a sound basis for interpreting the Balmer emission in the growing sample of magnetic O-stars being discovered in the MiMeS project \citep{Wad2012,Pet2012},
the present paper applies this concept of a dynamical magnetosphere to analyzing the rotationally modulated H$\alpha$ line emission of the prototypical magnetic O-star  \t1oc, but now using, for the first time, fully 
3D MHD simulations of the magnetic channeling and confinement of its radiatively driven stellar wind.

This extension to 3D represents a significant advance over the \citet{Sun2012} approach, which was based on simulations
that  assume 2D  axisymmetry, and so artificially suppress any variations in the azimuthal coordinate $\phi$.
To mimic the expected azimuthal fragmentation of structure in a 3D model, \citet{Sun2012} computed spectra from multiple snapshots of a 2D MHD simulation with large stochastic variations, which were then time-averaged as a proxy for the radiation-transport spatial averaging over azimuth in a full 3D model.

The 3D simulations here compute this lateral structure directly, applying the results to a full 3D radiative transfer model of the Balmer emission, and its variability with stochastic fluctuations of the wind, as well as with  the changing observer perspective over rotational phase.
The next section (\S 2) describes the basic MHD method, along with the parameters, boundary conditions, and numerical grid used.
This leads (\S 3) to a general description of the resulting 3D structure and evolution, with emphasis  on 
the  dense,  near-equatorial, magnetically confined gas that makes up the dynamical magnetosphere.
These simulation results are then applied (\S 4) within full 3D radiative transfer calculations to derive the Balmer line emission and its variation.
We conclude (\S 5) with a summary discussion of the implications for understanding the circumstellar emission of magnetic massive stars, along with an outline for future work.

\section{3D Simulation Method}

\subsection{Basic formalism}

For our  3D dynamical models we follow the basic methods and formalism developed for  2D simulations by \citet{udDOwo2002}, 
as generalized by \citet{Gag2005} to include a detailed energy equation with optically thin radiative cooling  \citep{MacBai1981}.
Since 3D modeling, especially MHD, is computationally intensive, we now use the parallel version of the 
publicly available numerical code ZEUS-3D called ZEUS-MP \citep{Hay2006}, which is an independent code but
employs essentially the same numerical algorithms, although in a parallel programming environment.

As in \citet{udDOwo2002},  the treatment of radiative driving by line-scattering follows the standard  \citet*[hereafter CAK]{Cas1975} formalism,
corrected for the finite cone angle of the star, using a spherical expansion approximation for the local flow gradients
\citep*{Pau1986,FriAbb1986}.
This  ignores {\em non-radial} components of the line-force  that may result from velocity gradients of the flow in the lateral directions. Although in purely hydrodynamic models these forces can have significant impact, e.g. by inhibiting
wind compressed disk \citep{OwoCra1996}, in magnetic models  they are generally small compared to the strong non-radial force components associated with the magnetic field \citep{udD2003}.  It  also ignores the extensive small-scale wind structure and clumping that can arise from the intrinsic instability of line driving \citep{Owo1988}, since modeling such dynamics requires a non-local line-force integration that is too computationally expensive to be tractable in the 3D MHD simulations described here.

\subsection{Numerical grid}

The computational grid and boundary conditions are again similar to those used by \citet{udDOwo2002}, 
as generalized by  \citet{udD2008} to include the non-zero azimuthal velocity at the lower boundary arising from rotation.
Flow variables are specified on a fixed 3D spatial mesh
in radius, co-latitude,  and azimuth: $\{r_i,\theta_j,\phi_k\}$.
The mesh in radius is defined from an initial zone $i=1$, which has a 
left interface at $r_1  = \Rstar$, 
the star's surface radius, out to a maximum zone ($i=n_r=300$), which 
has a right interface at 
$r_{300}  = 10 \Rstar$.
Near the stellar base, where the flow gradients are 
steepest, the radial grid has an initially fine spacing with
$\Delta r_{1}=0.0048 \Rstar$, and then increases by 1\% per zone out to a 
maximum of $\Delta r_{299}=0.095 \Rstar$.
This is fine enough to resolve gradients in the subsonic base, but is larger than used in previous models;
this allows for a larger time step, set as 0.3 of the Courant time,  giving a typical time step of 0.1~s.

The mesh in co-latitude uses $n_\theta=100$ zones to span the two 
hemispheres from one pole, where the $j=1$ zone has a left interface at
$\theta_1=0 \degr$, to the other pole, where the $j=n_\theta=100$ zone 
has a right interface at $\theta_{100}=180 \degr$.
To facilitate resolution of compressed flow structure near the magnetic
equator at $\theta= 90 \degr$, the zone
spacing has a minimum of $\Delta \theta_{49}= \Delta \theta_{50}=0.29 \degr $ about the equator, and 
then increases by 5\% per zone toward each pole, where 
$\Delta  \theta_1=\Delta  \theta_{99}=5.5 \degr $.

The mesh in azimuth is uniform, with $n_\phi = 120$ zones of fixed size $\Delta \phi = 3 \degr$ over the full 360$\degr$ range.
We also performed an initial  test run with twice the number of azimuthal zones for our future resolution study. But preliminary
results show that although there are some differences in finer structure during initial phases of breakup in azimuth, the asymptotic azimuthal structure scale is similar  to that found here.

\subsection{Boundary conditions}

Boundary conditions are implemented by specification of variables in
two phantom or ghost zones.
At the outer radius, the flow is invariably super-Afv\'enic outward, 
and so outer boundary conditions for all variables (i.e. density, energy,  and the
radial and latitudinal components of both the velocity and magnetic 
field) are set by simple {\it extrapolation}
assuming constant gradients.  On the other hand, at the inner radius, where the
surface of the star is defined, the flow is
sub-Afv\'enic. In general, the number of characteristics pointing into the grid determines
the number of boundary conditions needed to be specified \citep{Bog1997}. For supermagnetosonic flow,
this implies that all seven variables (density,  internal energy or pressure, three
components of velocity and two transverse components of the magnetic field) must be specified.
But for sub-magnetosonic flow here, only four characteristics point into the grid (entropy, fast, slow and Afv\'enic).
Thus, four variables must be specified and the remaining
three of the variables need to be extrapolated \citep{Bog1997}. These conditions are briefly described below.

In the two radial zones below $i=1$, we 
set the radial velocity $v_{r}$ by constant-slope extrapolation,
and 
fix the density at a value $\rho_o$
chosen to ensure subsonic base outflow for the characteristic
mass flux of a 1D, nonmagnetic CAK model, i.e.
$\rho_o \approx {\dot M}/(4 \pi \Rstar^2 a/6)$.
These conditions allow the mass flux and the 
radial velocity to adjust to whatever is appropriate for
the local overlying flow
(Owocki, Castor, and Rybicki 1988).
In most zones, this corresponds to a subsonic wind outflow,  although inflow 
at up to the sound speed is also allowed.

Magnetic flux is introduced through the lower boundary as the radial
component of a dipole field $B_r (\Rstar, \theta) = B_o \cos (\theta)$, where
the polar field $B_o = 1100$~G is adopted from the observationally inferred value for \t1oc \citep{Don2002}.
The latitudinal components of magnetic field, $B_\theta$, and flow speed $v_\theta$, are again  set by constant slope
extrapolation.
The azimuthal field $B_\phi$ is likewise set by extrapolation, while the azimuthal speed is set by the assumed stellar rotation,
$v_\phi(\Rstar, \theta ) = V_{rot} \sin \theta $.

While the code is fully 3D, the choice of spherical coordinates leads to numerical difficulties near the coordinate singularity along the polar axis, which is treated with reflecting boundary conditions.
Because this prohibits any flow across the axis, we ignore the slow, {\em oblique} rotation inferred for \t1oc, since at a period of $\sim$15~d, it is dynamically unimportant.
However, to avoid pockets of rarefied gas near the stellar surface that lead to very small Courant time, we do introduce a small dynamically insignificant {\em field-aligned} rotation with an equatorial surface speed $V_{rot} = 10 \, \kms$.

\subsection{Stellar parameters and initial condition}

The adopted stellar and  wind parameters are summarized in Table~1.
Using these parameter values, we first relax a non-magnetic, spherically symmetric wind model to an asymptotic steady state; this yields the  quoted mass loss rate and terminal speed, which are in very good agreement with the standard CAK scaling,  adjusted for a finite-disk and non-zero sound speed (Owocki \& ud-Doula 2004).
This relaxed CAK steady state is used for the initial condition in density and velocity.
The temperature is initially set to the associated stellar effective temperature, $T_{\rm eff} \approx 40,000$~K, which also subsequently sets a floor for the temperature, to approximate a radiative equilibrium in which photoionization heating counters the cooling by line emission and adiabatic expansion.

The magnetic field is initialized to have a simple dipole form with components 
$B_r=B_o (\Rstar / r)^3 \cos \theta$, 
$B_\theta= (B_o/2) (\Rstar / r)^3 \sin \theta$, 
and $B_\phi = 0$,
with $B_o$ the polar field strength at the stellar surface.
From this initial condition, the numerical model is then evolved 
forward in time to study the dynamical competition between the field and flow, as detailed in \S 3.

\begin{table}
\caption{Stellar and wind parameters}
\begin{tabular}{|l|l|l|}
\hline
\hline
Name&Parameter&Value\\
\hline
Mass		&$M_\ast$	&$40 \Msun$\\
Radius		&$\Rstar$		&$8 \Rsun$\\
Luminosity	&$\Lstar$		&$1.5 \times 10^5 \Lsun$\\
CAK $\alpha$	&$\alpha$		&0.5\\
Line strength	&$\bar{Q}$	&700\\
Mass loss	rate	&$\dot M $	&$ 3.3 \times 10^{-7} \Msun$yr$^{-1}$\\
Terminal speed&$\vinf$	         & $3200$ km s$^{-1}$\\
Polar field		&$B_{Pole}$	&$1100$~G\\
\hline
\end{tabular}
\label{tab1}
\end{table}%

Following \citet{udDOwo2002}, the stellar, wind, and magnetic parameters in Table~1 can be used\footnote{We emphasize here that mass loss rate and flow speed used to estimate this magnetic confinement parameter (and thus the associated Alfv\'{e}n radius) should {\em not}, as is sometimes done, be based on any observationally inferred values using {\em non-magnetic} analysis for the star in question, as the magnetic field may significantly influence the predicted circumstellar density and velocity
structure. Rather, $\eta_\ast$ is defined using the mass loss and speed {\em  expected} for a corresponding {\em non-magnetic} star of that spectral and luminosity class,  as inferred from scaling laws \citep[e.g.,][]{Vin2001} based on line-driven-wind theory.}  to define a dimensionless ``wind magnetic confinement parameter'',
\beq
\eta_\ast \equiv \frac{B_{eq}^2 \Rstar^2}{{\dot M} \vinf} \approx 14 
\, ,
\label{etastar}
\eeq
where for a dipole, the equatorial field is just half the polar value, $B_{eq} = B_o/2$.
Since $\eta_\ast > 1$, the magnetic field dominates the wind outflow near the stellar surface.
But because the magnetic energy density has a much steeper radial decline ($ B^2 \sim r^{-6}$ for a dipole) than the wind ($\rho v^2 \sim r^{-2}$ if near terminal speed), the wind can strip open field lines above a
a characteristic {\em Alfv\'{e}n radius}, set approximately by  \citep{udD2008}
\beq
\frac{R_{\rm A}}{\Rstar} 
\approx 0.3 + \eta_\ast^{1/4}
 \approx 2.23
\, .
\label{radef}
\eeq

\begin{figure}
\begin{minipage}{0.9\linewidth}
\centering
\includegraphics [width=\textwidth, angle=0]{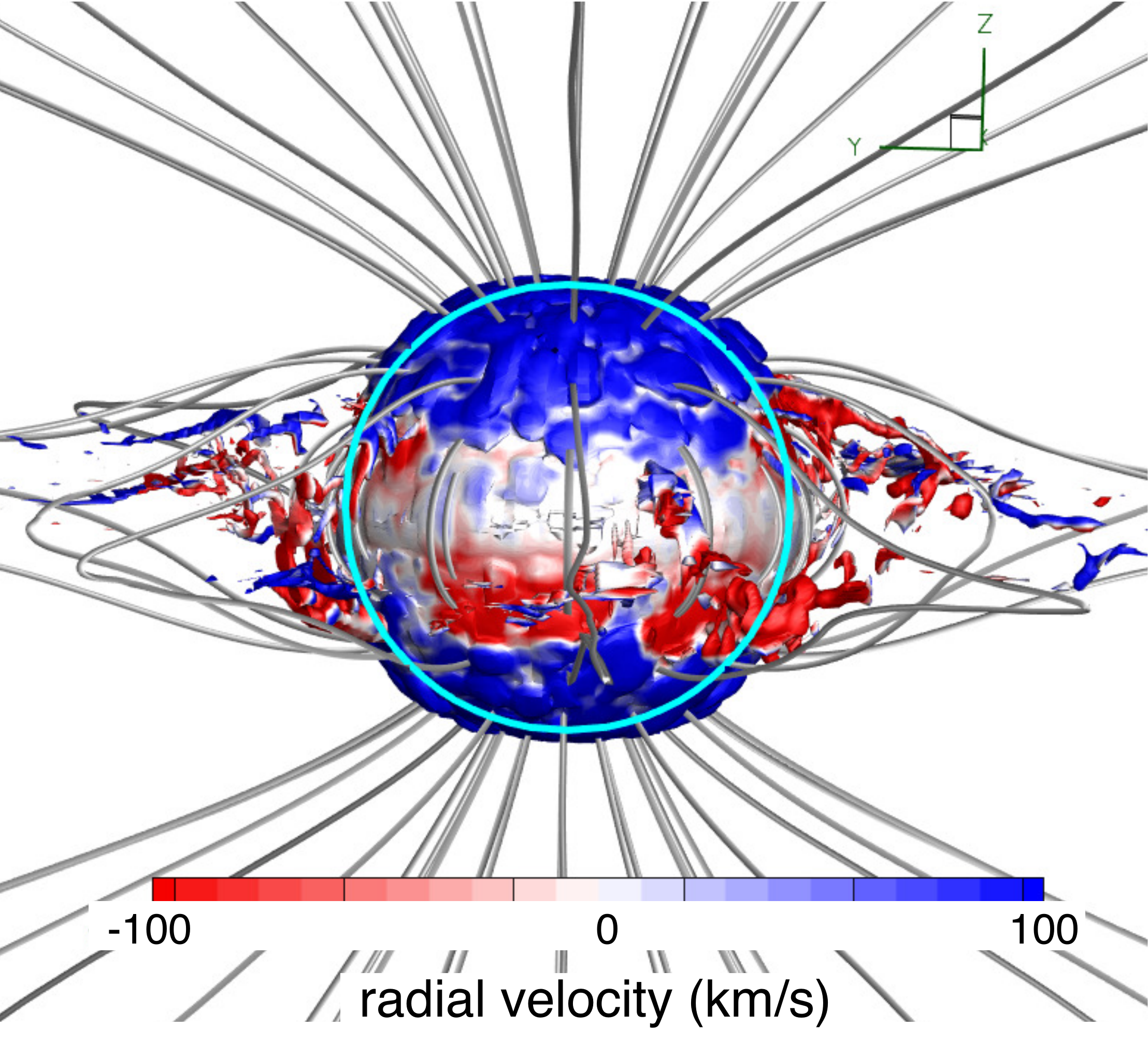}
\end{minipage}
\caption
{ A near stellar surface view of an iso-density ($\log\rho=-12.5$ in {\em cgs} units) surface, colored with radial velocity $v_r$, 
ranging from $-100 \, \kms$ (red) to $+100 \, \kms$ (blue).
 Note how some of the gas material within the closed loop region falls back onto the stellar surface, while 
 material in the open field (polar) region streams radially outward. The slowly moving (or infalling) region depicted as white (red) 
 color indicates the lateral extent of the magnetosphere. The circle (cyan) represents the stellar surface.
 }
\label{GlobalStructure}
\end{figure}

\section{MHD Simulation Results}

\subsection{Snapshot of 3D structure}
\label{global}

As background for the Balmer emission line computations in \S 4, let us here characterize the spatial structure and time evolution resulting from our 3D MHD simulation.

The characteristic time for the CAK wind here to flow from the sonic point near the stellar surface to the outer boundary at $10 \Rstar$ is about $t_{flow} \approx 25$~ks. For MHD simulations we find in practice \citep{udDOwo2002} it takes  $\sim 10-15$ such flow times for resulting structure to reach a state that is independent of the initial conditions. 
To provide an extended sampling of random variations in this asymptotic state, we run the 3D simulations here to a final time $t=1000$~ks, corresponding to roughly 40 wind flow times.

Figure \ref{GlobalStructure} shows a close-in, equatorial view of the dichotomy between  open field over the poles and closed loops around the equator, with density rendered as an iso-density surface for $\log \rho = -12.5$~g/cm$^{3}$.
The color represents the associated {\em radial} (in grid coordinate system) component of the flow velocity, ranging from -100~$\kms$ inflow (red) to +100~$\kms$ outflow (blue).
Note that the polar regions have mainly just outflow, reflecting the open nature of the field.
In contrast, the equatorial regions have a complex combination of outflow and inflow, reflecting the trapping of the underlying base wind within closed loops, with subsequent gravitational fall-back onto the stellar surface. 

A similar pattern of wind trapping and subsequent infall was also seen in 2D simulations \citep{udDOwo2002, udD2008}, manifested there by a complex `snake' pattern extending over radius and latitude, but formally coherent in azimuth within the imposed 2D axisymmetry.
This 3D simulation now shows how,  because of spontaneous symmetry breaking,  this infall becomes azimuthally fragmented into multiple, dense, infalling clumps, distributed in both latitude and azimuth on a scale that is comparable to the latitudinal scale of the 2D snakes. 
 
\begin{figure*}
\begin{center}
\includegraphics [scale=0.70]{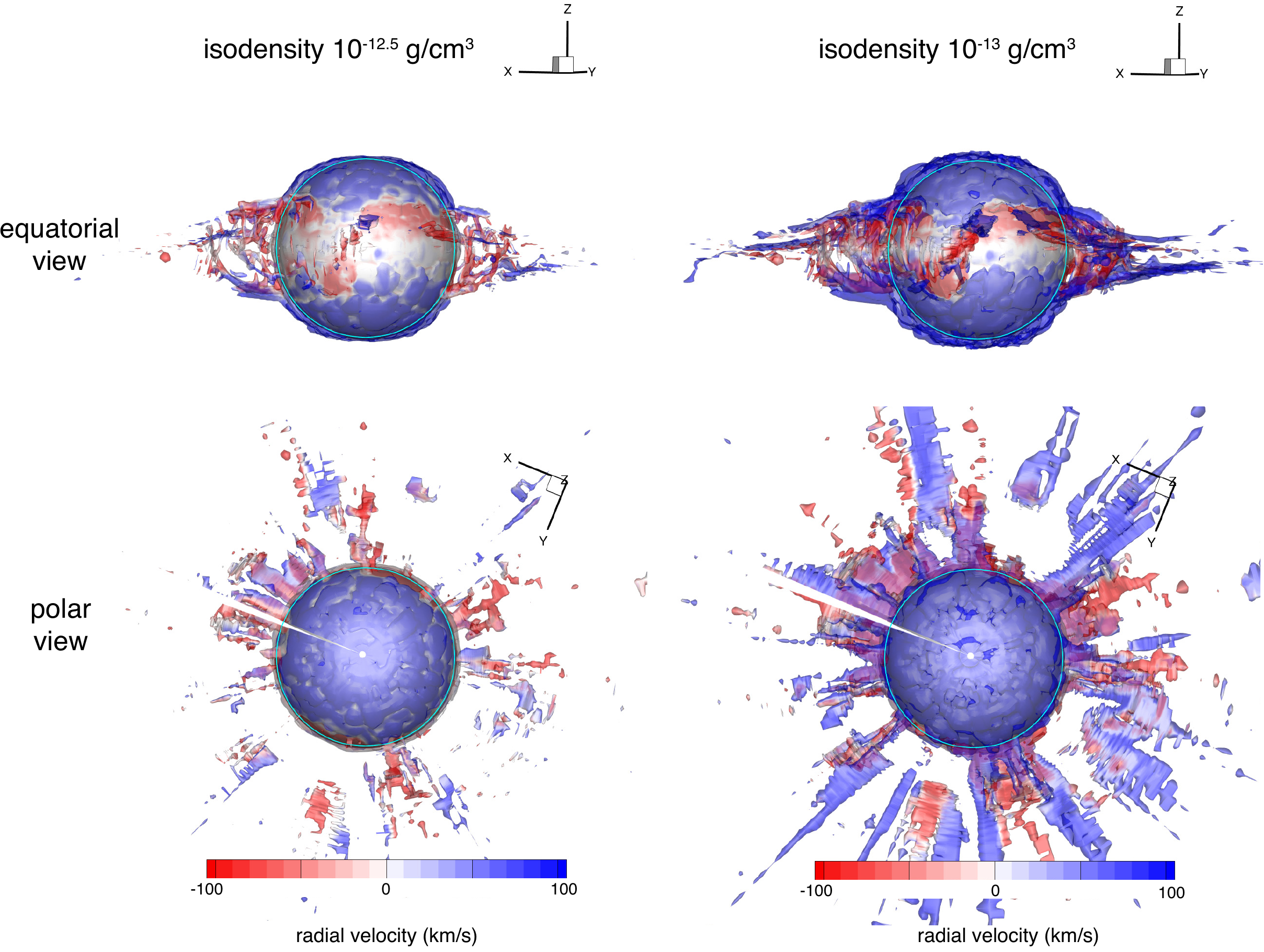}
\end{center}
\caption
{Iso-density surfaces at log($\rho$)=-12.5 (left) and log($\rho$)=-13.0 (right) for a fixed time snaphot $t=1$~Ms, viewed from the equator (top panels)
or pole-on (bottom panels). As in figure \ref{GlobalStructure}, the color represents radial velocity. The equatorial `belt' visible as white band (low velocity) represents the latitudinal extent of the magnetosphere. Note that substantial material within that region is falling back onto the stellar surface, thus forming the  dynamical magnetosphere. 
The cyan circle again represents the stellar surface.
}
\label{isorho}
\end{figure*}

To provide a more complete picture of this complex, 3D wind structure, 
figure \ref{isorho} shows both equatorial (top) and polar (bottom) views for two values of the iso-density surface, namely $\log \rho = -12.5$ (left) and $\log \rho = -13$ (right), again with the same color coding for radial velocity.
The polar view now shows clearly the azimuthal organization and scale of the trapped equatorial structure, and how it becomes extended outward into radial ``spokes''. 
While complex, the lower iso-density plot on the right shows a general trend from dominance of infall in the inner region (marked by red) to outflow at larger radii (marked by blue).

Overall, this represents a 3D generalization of the dynamical magnetosphere concept developed to characterize 2D simulations, and their application to modeling Balmer line emission 
\citep{Sun2012}.
The essential feature is that the trapping and compression of wind material in closed loops results in a substantial enhancement of the time-averaged density near the magnetospheric equator, leading then to an associated enhancement in line emission, as we  demonstrate quantitatively in \S 4.

\subsection{Time development and evolution of wind structure}

Let us next consider how such complex spatial structure develops from the steady, spherical initial condition, and how it subsequently evolves into a stochastic variation associated with wind trapping, breakout, and infall.
Since much of the key structure is near the magnetic equator, let us follow the approach developed by \citet{udD2008} that defines a radial mass distribution $dm_{\rm e}/dr$ within a specified co-latitude range $\Delta \theta $ (taken here to be  $10 \degr$)  about the equator,
 \beq
\frac{dm_{\rm{e}} (r,\phi,t)}{dr} \equiv 2 \pi r^{2}
\int_{\pi/2-\Delta \theta/2}^{\pi/2+\Delta \theta/2} \, \rho(r,\theta,\phi,t) \,
\sin \theta \, d\theta 
\, .
\label{dmedrDef}
\eeq
In previous 2D models, the imposed axisymmetry means this $dm_{\rm e}/dr$ is a function of just radius and time, and so can be conveniently rendered as a colorplot to show the time and spatial variation of the equatorial material.
In the present 3D case, there is an added dependence on the azimuthal coordinate $\phi$, but upon averaging over this, we again obtain a function of just radius and time,
\beq
\frac{d{\bar m}_{\rm{e}} (r,t)}{dr} \equiv \frac{1}{2 \pi} \int_0^{2\pi} \frac{dm_{\rm e}}{dr} (r,\phi,t) \, d\phi
\, .
\label{dmdrphiavg}
\eeq

\begin{figure}
\includegraphics[width=83mm]{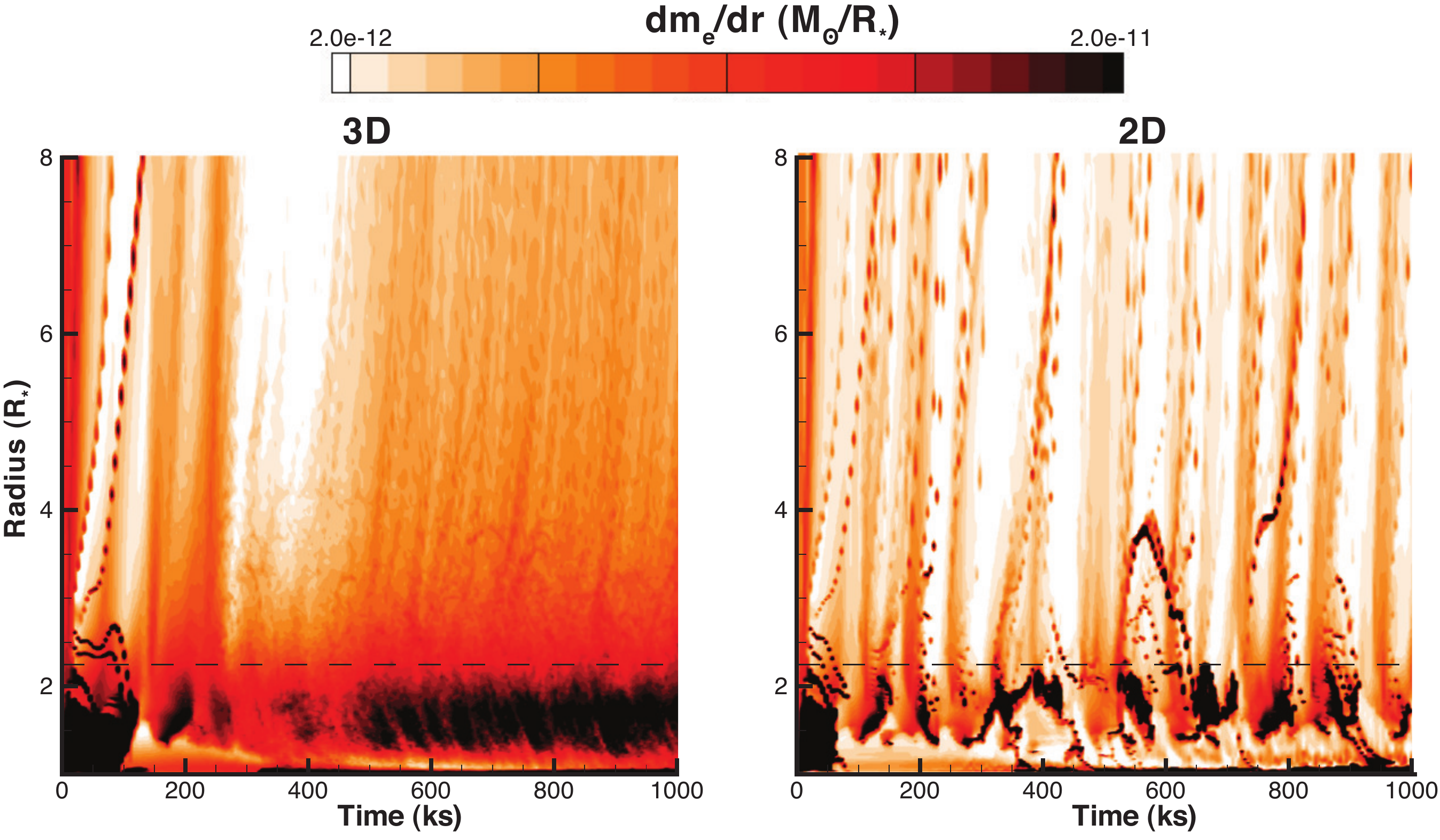}
\caption
{Equatorial mass distribution $dm_{\rm e}/dr$ (in units $M_\odot/R_\ast$) for the azimuthally averaged 3D model (left) and for a corresponding 2D model (right), plotted versus radius and time, with the dashed horizontal line showing the Alfv\`en radius $R_{\rm A} \approx 2.23 \Rstar$.
Following a similar adjustment to the initial condition,  the long-term evolution of the 2D models is characterized by a complex  pattern of repeated strings of infall, whereas the 3D azimuthally averaged model settles into a relatively smooth asymptotic state characterized by an enhanced mass near and below the Alfv\'{e}n radius.
}
\label{dmdr-r-vs-t}
\end{figure}

Figure \ref{dmdr-r-vs-t} compares the time and radius variation of the $d{\bar m}_{\rm e}/dr$ from the 3D simulation (left) with the corresponding $dm_{\rm e}/dr$ from an analogous 2D model  (right) with identical parameters.
The dashed horizontal  line, marking the estimated Alfv\'{e}n radius $R_{\rm A} \approx 2.23 \Rstar$, roughly separates the regions of outflow\footnote{Actually, for the 2D case,  the separation is more distinct in isothermal models shown by \citet{udD2008}; in the 2D simulations here, the inclusion of a full energy balance leads to episodes of field reconnection that pull relatively high-lying material back to the surface. The reasons for these subtle differences will be investigated in a follow-up study.} along open field lines above $R_{\rm A}$, from the complex pattern of upflow and downflow for material trapped in closed magnetic loops below $R_{\rm A}$.
In the 2D models, the complex infall pattern occurs in distinct, snakelike structures, but in the 3D models, these break up into multiple azimuthal fragments that, upon averaging, give the 3D panel a smoother, more uniform variation.
More importantly, note the clear enhancement in the mass just below the Alfv\'{e}n radius, representing what we characterize here as the dynamical magnetosphere.
 
\begin{figure}
\centering
\includegraphics [width=8.75cm,angle=0]{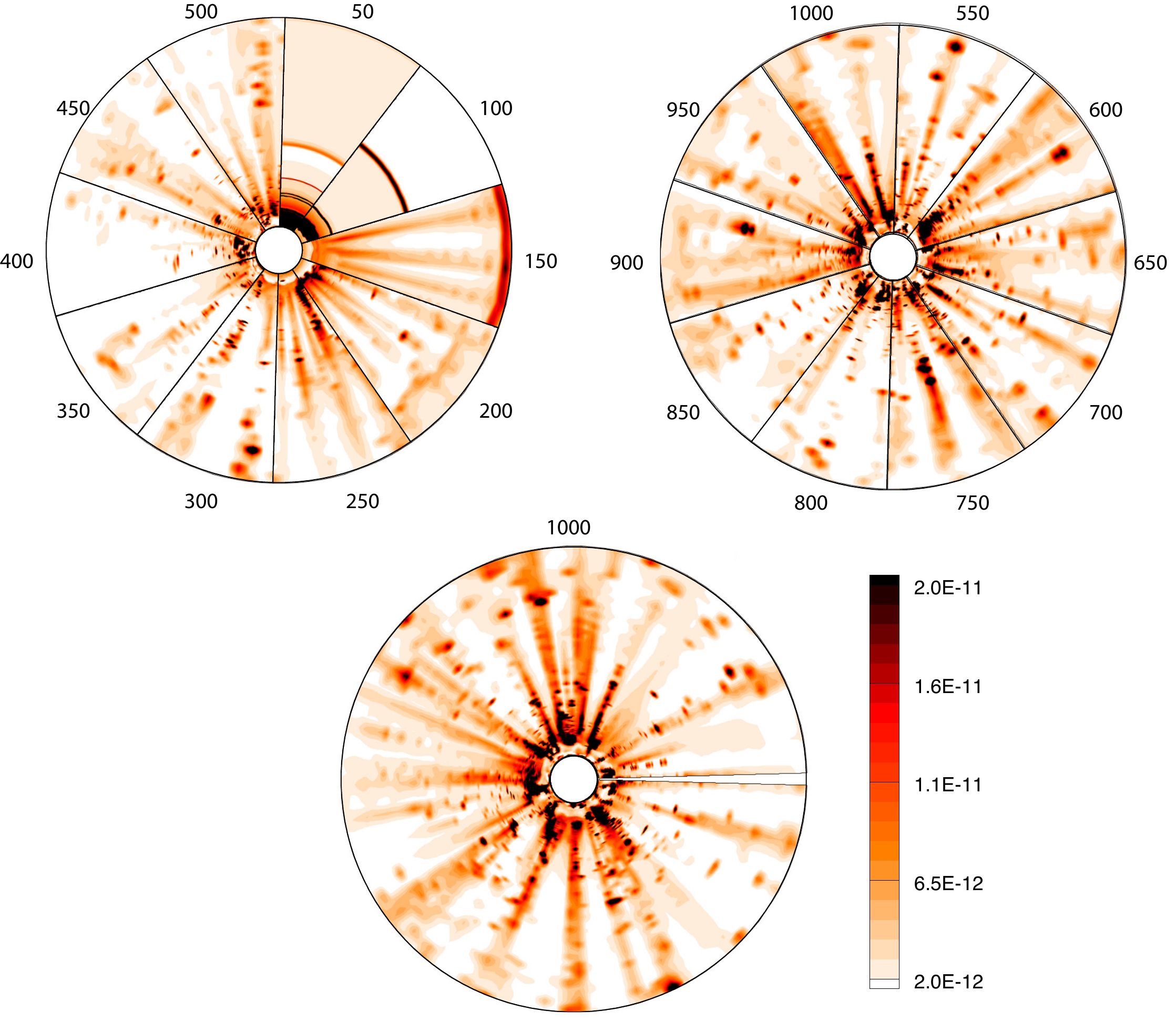}
\caption
{Time evolution of $dm_{\rm e}/dr$ for an arbitrary $36 \,\degr$ azimuthal segment of the 3D model, at marked time intervals of 50~ks ranging over $t=50-500$~ks in the upper left panel, extending to $t=550-1000$~ks in the upper right.
The bottom panel shows the full variation over $360 \, \degr$ for the final simulation time $t=1$~Ms.
As in figure \ref{dmdr-r-vs-t}, the units of $dm_{\rm e}/dr$ are $\Msun/\Rstar$.
}
\label{dmdr-pie}
\end{figure}

To illustrate the time evolution of the azimuthal breakup and variation in 3D wind structure, the upper panels of figure \ref{dmdr-pie} use a ``clock format'' to show the time progression of an arbitrary 36$\degr$ segment of $dm_{\rm e}/dr (r,\phi,t)$ at time steps of 50~ks, ranging over $t=50-500$~ks for the left panel, and continuing through $t=550-1000$~ks in the right panel.
Following the spherically symmetric initial condition, the magnetic field first channels the wind into a compressed, azimuthally symmetric,  disk-like structure ($t=50 ~ \& ~100$~ks), which however then breaks up (by $t=150$~ks) to a series of radial spokes with an azimuthal separation of roughly $\Delta \phi \approx 15-20 \degr$. 
Remarkably, following this initial breakup, the azimuthal structure keeps the same overall character and scale within a complex stochastic variation\footnote{Initial test runs done at twice the azimuthal grid resolution from that used here show an initially smaller scale for the breakup, which however subsequently (by $\sim 300$~ks) settles into a scale that is similar to what is found here. Details will be given in a future paper that carries out a systematic resolution study for such 3D MHD simulations.}. 

Note in particular that the structure over the full $360 \, \degr$, as shown for the final time $t=1000$~ks in the lower row of figure \ref{dmdr-pie}, is quite similar to that shown for the evolving segments in the right-side upper row for $t = 550 - 1000$~ks.
Moreover, the $\sim 500$~ks lower bound for this settling time corresponds to the time in figure \ref{dmdr-r-vs-t} for the formation of  the enhanced $dm_{\rm e}/dr$ that characterizes the dynamical magnetosphere.

As noted in \S \ref{global}, the inferred azimuthal scale of $15-20 \degr$ is comparable to the scale of infalling ``snakes'' found in 2D simulations, which typically extend over some fraction of the loop length between closed footpoints. For the present case, the maximum latitude of a closed loop is about $45 \degr$, giving a typical angle extent of about $15-20 \degr$ for both the snakes and the azimuthal breakup. Future studies should examine whether these structure scales do in fact follow the expected loop closure latitude, which should scale with magnetic confinement parameter as $\arcsin {\eta_\ast^{-1/8}}$.

In summary, the equatorial structures seen in previous 2D models now become azimuthally fragmented with a characteristic separation of $\Delta \phi \approx 15-20 \degr$, which however upon averaging in azimuth, have a radial and latitudinal variation that is similar to what is obtained from {\em time}-averaging in 2D models. 
This has important implications for modeling the associated Balmer line emission, as we discuss next.

\section{Balmer line emission: model vs. observations}
\label{Halpha}

Now let us  examine to what extent the 3D MHD wind simulations
presented here can reproduce the observed H$\alpha$ emission and
variability in \t1oc.

Applying the 3D radiative transfer method developed by \citet{Sun2012} to the 3D MHD simulations described above,
we compute synthetic H$\alpha$ profiles by solving the formal
integral of radiative transfer in a cylindrical coordinate system
aligned toward the observer. The angle $\alpha$ defines the observer's
viewing angle with respect to the magnetic pole, and, for given
inclination $i$ and obliquity $\beta$,
\begin{equation} 
  \cos \alpha = \sin i \sin \beta \cos \Phi + \cos i \cos \beta
  \label{Eq:trig}
\end{equation}
gives the observer's viewing angle as a function of rotation phase
$\Phi$. Due to the slow rotation of the targeted star, we may neglect
any dynamical effects of rotation and use the same simulation for all
obliquity angles $\beta$. The computations further use a pre-specified
photospheric line profile as a lower boundary condition, and solve the
equations of radiative transfer only in the circumstellar structure. Hydrogen NLTE
departure coefficients are estimated from a spherically symmetric
unified model atmosphere calculation (using {\sc fastwind}, \citealt{Pul2005}). Moreover, since the energy equation in the MHD
simulation gives only a rough estimate of the wind temperature
structure, we also take this from the {\sc fastwind} model, except for
in shock-heated regions with $T > 10^5$, where we set the H$\alpha$
source function and opacities to zero. For further discussion about
these and other assumptions entering the Balmer line profile
calculations, see \citet{Sun2012} and references therein.

\subsection{Periodic rotational phase variability}

\begin{figure}
\resizebox{\hsize}{!}{\includegraphics[angle=0]{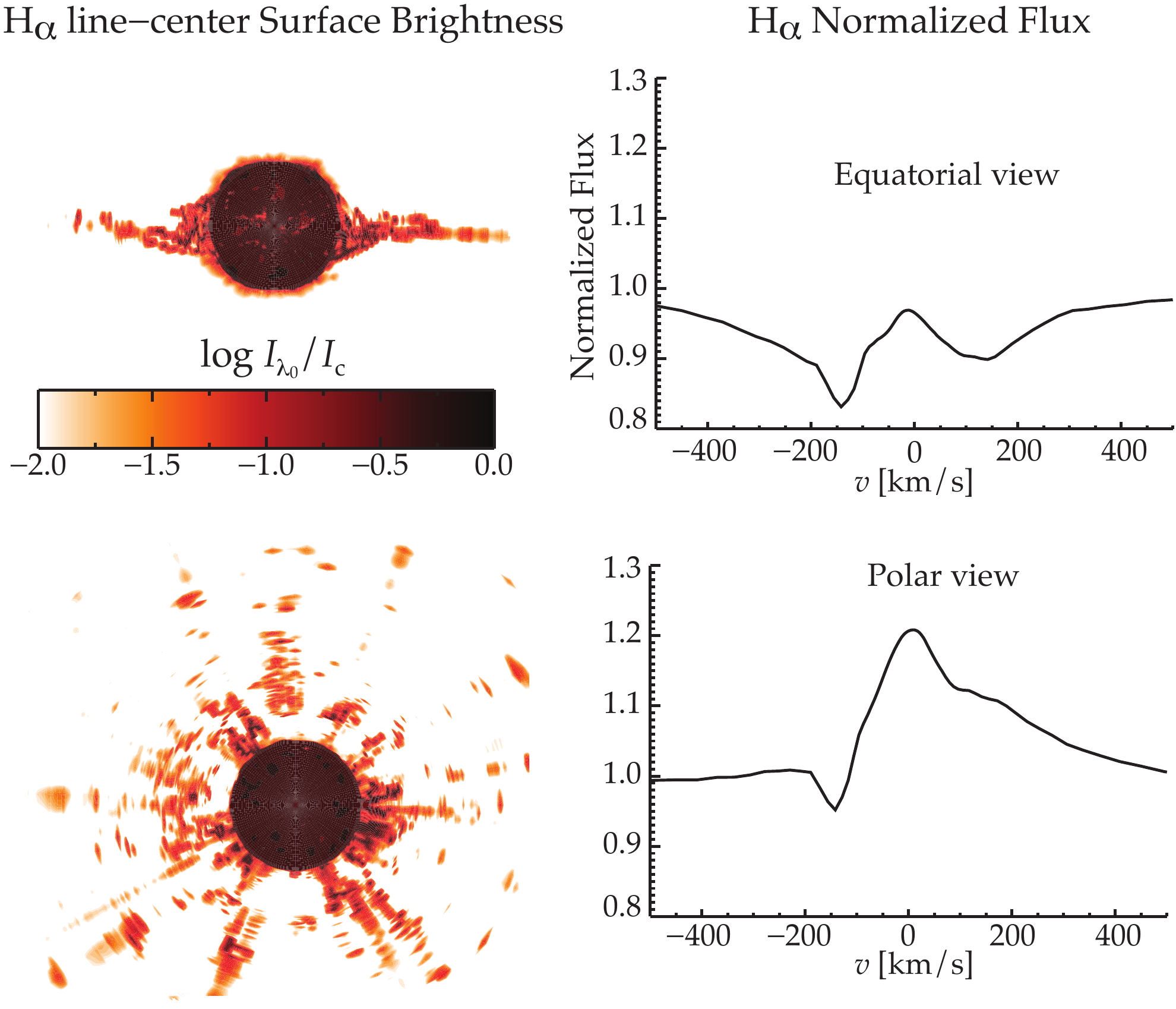}}
\caption{Synthetic H$\alpha$ surface brightness maps (left panels) and
  flux profiles (right panels) for observer viewing angles above the
 equator (upper panels) and magnetic pole (lower panels),  as
  calculated from a snapshot of the 3D MHD wind simulation taken at 1000
  ksec after initialization.}
\label{Fig:hasb}
\end{figure}
\begin{figure}
\resizebox{\hsize}{!}{\includegraphics[angle=0]{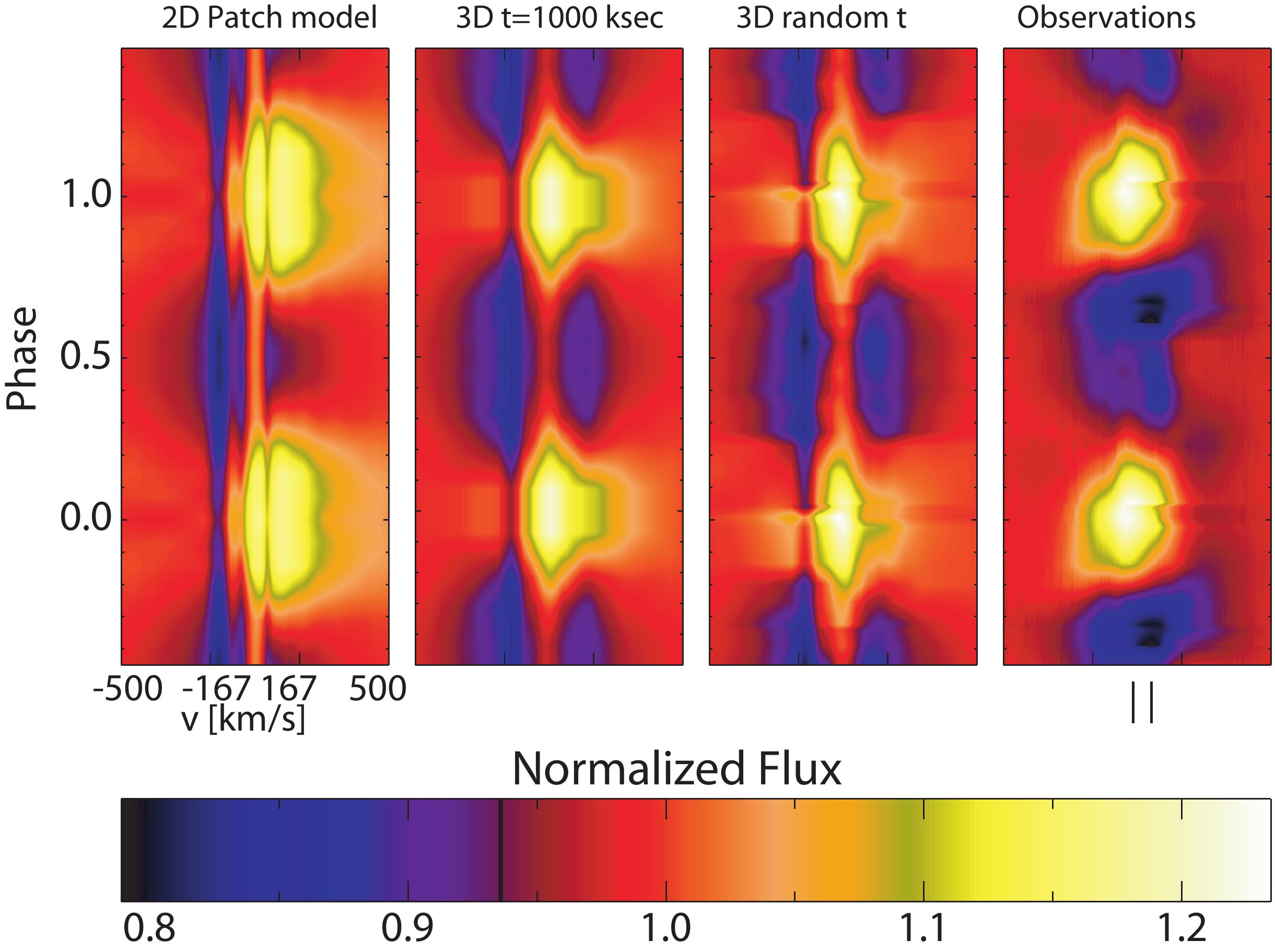}}
\caption{Observed and synthetic H$\alpha$ dynamic spectra of
  $\theta^1$ Ori C. Synthetic spectra calculated from patching
  together 2D model snapshots (panel 1), a single 3D model snapshot
  (panel 2), and from a random selection of 3D snapshots (panel 3).
  The vertical small lines
  in panel 4 represent the velocity range of the nebular emission.
  }
\label{Fig:hadyn}
\end{figure}

Figure \ref{Fig:hasb} displays synthetic H$\alpha$ line-center surface
brightness maps and flux profiles viewed from above the
magnetic pole and equator, corresponding approximately to rotational phases of 0.0 and 0.5, respectively, according to the ephemeris of \citet{Sta2008}. As in \citet{Sun2012}, we
interpret the H$\alpha$ variability within the framework of a dynamical
magnetosphere, but now using the full 3D dynamical model for the azimuthal wind structure.
For the pole-on view, the surface brightness of line-center emission in the lower panel of figure \ref{Fig:hasb} follows closely
 the lateral fragmentation seen from the corresponding iso-density plots in the bottom row of figure \ref{isorho}.
 Similarly, there is a good correspondence for the equatorial view, but now the projection leads to an overall lower level of the total emission, associated with the smaller projected surface area of the structure.
 As shown by the right panels of figure \ref{Fig:hasb}, this leads to a stronger level of emission in the line profiles for the polar view.
 But note that even for the equatorial view, the circumstellar emission still partially refills the photospheric absorption profile. 
 Most of the emission arises from the relatively dense material trapped in the closed loops near the stellar surface, with only about 10-20\% arising from the extended radial spokes. Overall, the magnetospheric structures found here have quite different morphology from compressive structures generated by instabilities of line-driving in non-magnetic stars \citep{Fel1997,DesOwo2003}. 
 
Within the context of the MiMeS project, 18 high-resolution, high-SNR observations of \t1oc were obtained with the spectropolarimeter ESPaDOnS at the Canada-France-Hawaii telescope (4 from \citet{Pet2008}  and 14 from MiMeS) which were used to produce the dynamical spectra of H$\alpha$  shown in figure~\ref{Fig:hadyn}. Leveraging the high spectral resolution (4 km\,s$^{-1}$), we remove the strong nebular line by fitting a polynomial curve to the adjacent sides of the line profile. This procedure provides us with an indication of how smoothly these lines vary in phase.
Nonetheless, localised features within this velocity range (illustrated by small vertical lines in the fourth panel in figure~\ref{Fig:hadyn}) could still be hidden.

Figure \ref{Fig:hadyn} compares observed dynamic H$\alpha$ spectra of 
$\theta^1$ Ori C over its rotational phase (far right) with 3 models that each assume
 $i = \beta = 45 \degr $, but which have increasing levels of sophistication progressing from left to right.
 
 The first panel is
computed from 100 randomly selected 2D simulation snapshots that have
been patched together to form an `orange slice' pseudo-3D model with
an azimuthal coherence of 3.6$\degr$. This patch approach is similar to
that used to reproduce periodic H$\alpha$ variability in
the magnetic O-stars HD\,191612 and HD\,57682 from 2D MHD simulations
\citep{Sun2012, Gru2012}. 

The second panel then
displays spectra computed from a single snapshot of the 3D simulation,
and illustrates that such full 3D model spectra are in fact
qualitatively very similar to the averaged 2D models, as was suggested by
\citet{Sun2012}.

Finally, the third panel in
Figure ~\ref{Fig:hadyn} is constructed from a random selection of 3D
snapshots with phases that coincide with the observations seen in the fourth panel; it is discussed further in the following subsection.

 Overall, all three models give reasonable and comparable agreement with the observed dynamic spectrum.  This suggests that one can indeed use the more simplified approaches to explore how observational diagnostics constrain the properties of a dynamical magnetosphere, calibrated with  selected comparisons with more computationally demanding, and physically realistic  3D models.

Although the simulations reproduce quite well both the magnitude and phase of the observed rotational
variation, the observed profiles exhibit an
asymmetry about phase 0.5 that is not found in our numerical models;
such asymmetries may
have their origin in, e.g., non-dipolar components of the magnetic
field, consideration of which we defer to future study.
Moreover, to match the absolute level of emission, we
have increased the H$\alpha$ scaling invariant $ \sqrt{f_{\rm cl}}
\dot{M} / (v_\infty R_\star)^{3/2}$ of the
underlying simulations by a factor of 2.5,
where $f_{\rm cl}$ is an overdensity correction factor that accounts for small-scale clump structure \citep[e.g.,][]{Pul2008}.
The need for this increase
may indicate the MHD simulations either assume a somewhat too low
mass-loss rate for $\theta^1$ Ori C, or underestimate the
overdensities of the emitting H$\alpha$ material around the magnetic
equator.

\subsection{Stochastic short time-scale variability}

\begin{figure}
\resizebox{\hsize}{!}{\includegraphics[angle=90]{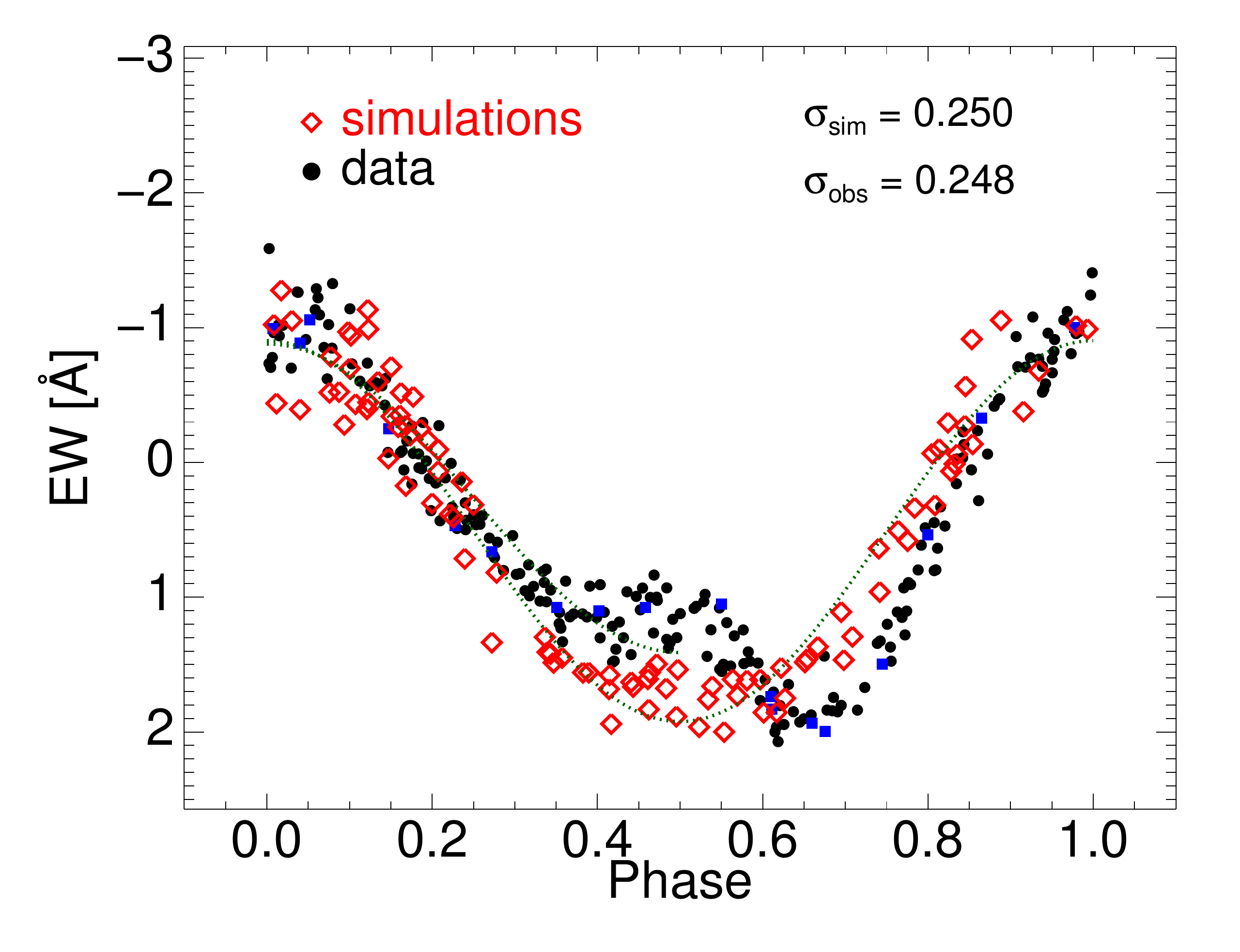}}
\caption{Observed and synthetic H$\alpha$ equivalent-width
  curves. The synthetic measurements are calculated from 100 random 3D
  simulation snapshots phased randomly. The quoted $\sigma$'s measure
  standard deviations from functional fits (dotted green lines) to
  simulations and data. 
  The blue squares denote the additional observational data points that correspond to the dynamical spectra in figure~\ref{Fig:hadyn}.
  See text.}
\label{Fig:haew}
\end{figure}

The H$\alpha$ emission of $\theta^1$\,Ori\,C has been monitored regularly for nearly two decades and an extensive set of equivalent width measurements has been compiled by \citet{Sta2008}
shown as the black dots in figure \ref{Fig:haew}, phased according to the 15.4 d rotation period using the ephemeris   of \citet{Sta2008}.
Although the level of both emission and its modulation with rotational phase are very stable,  there is some significant random variability present during all phases.
 Such stochastic H$\alpha$ scatter may originate from small differences in the amount of material trapped in the dynamical magnetosphere at any given time.

To examine the level of such short time-scale variability in our 3D
MHD model, we randomly select simulation snapshots
between 800-1000 ksec and from these calculate line profiles for
random rotational phases. 
The results show that the model closely reproduces  both the {\em systematic} variation with rotational phase, 
as well as the {\em stochastic} deviation in the multi-year data from the mean at each phase.
The quoted standard deviations $\sigma_{\rm sim} \approx \sigma_{\rm obs} \approx 0.25
\, \rm \AA$ have been estimated by calculating the variance from separate
functional fits $f(\alpha) = a + b|\cos \alpha|$ to the model and observational data.
(Due to the asymmetry, data are fit only between phases 0 and 0.5.)
Though the interpretation of this simple estimate is  complicated by the fact that  non-magnetic O-stars with
significant wind strength also show stochastic H$\alpha$ variability \citep{Mar2005}, it nevertheless suggests that in \t1oc the observed H$\alpha$ scatter is 
likely dominated by processes associated with the star's dynamical magnetosphere.

This approach of randomizing variations in model equivalent width can also be used to account for the effect of stochastic wind variations on the synthetic dynamic spectrum, as illustrated in the third panel of  figure~\ref{Fig:hadyn}. 
We have aligned synthetic model phases to those for the subset of times with available spectra from our MiMeS observations (blue squares in figure \ref{Fig:haew}).
Comparison of the third and fourth panels of figure~\ref{Fig:hadyn} again show an overall good agreement between the models and observations.

\section {Discussion, Conclusions and Future Work}

A central advance of the work described here is the incorporation of all three spatial dimensions in the MHD simulation model. This allows us to follow directly the spontaneous breaking of the azimuthal symmetry that was imposed in previous 2D models, while still confirming a global structure that is quite similar to that found in those models. At high latitude this is characterized by fast wind outflow along open field lines. At low latitudes,  an equatorial belt of closed loops traps the wind  into a complex pattern of upflow and infall near the star, but this gives way again to open field outflow above the Alfv\'{e}n radius. 

The addition of a third dimension along the azimuth means that this complex pattern now fragments into distinct regions of upflowing wind and infalling clumps. This leads to an associated fragmentation of the wind breakout above the Alfv\'{e}n radius, resulting in an alternating pattern of radial spokes of slow, high-density outflow, with a characteristic azimuthal separation of $15-20 \degr$, divided roughly equally between the dense spokes and a more rarefied flow between them. Such a separation angle is well above the adopted azimuthal grid size $\Delta \phi \approx 3 \degr$, and indeed a higher resolution test run with half this grid size gives asymptotic wind structure at a similar scale. As such, this scale does not appear to be strongly dependent on the grid resolution, though further investigation will be needed to test this and clarify what physical processes set the structure size.

An additional focus here is the application of these 3D MHD simulation results to modeling the H$\alpha$ line emission from the prototypical magnetic massive star \t1oc, representing a 3D extension of the dynamical magnetosphere paradigm recently introduced by \citet{Sun2012}.
As a proxy for the spatial averaging over the azimuthal variations expected from a full 3D model,  this previous work computed time-averaged spectra from a strictly 2D axisymmetric simulations with  stochastically varying wind structure.
As demonstrated by the close similarity of the 2D vs. 3D models for the dynamic spectra in figure \ref{Fig:hadyn}, a key result here is to confirm a general validity for this proxy approach. 
This suggests then that future modeling of such rotational variation of line emission in the growing number of slowly rotating magnetic O-type stars could be achieved with reasonable accuracy using much less computationally expensive 2D MHD simulations to characterize the dynamical magnetosphere.
Of course, such an approach is only possible for the simple, axisymmetric field configuration of a magnetic dipole.
To account for deviations from a dipole, such as might be inferred from the asymmetric phase variation in the  equivalent width light curve here, one will again need to carry out fully 3D simulations.

Apart from this modest non-dipole asymmetry,  both the 2D and 3D dipole-based models here match quite well the rotational phase variation of H$\alpha$.
But a further quite remarkable result of the present study is how well the synthetic 3D, ``random time" model (third panel in figure \ref{Fig:hadyn}) reproduces also the observed {\em stochastic} variation in the H$\alpha$ equivalent width, as demonstrated  in figure \ref{Fig:haew}.
This suggests that the number and sizes of distinct structures generated in the simulated dynamical magnetosphere may be quite comparable to that in the actual star.

On the other hand, the  magnetic O-star HD~191612, which has a stronger magnetic confinement ($\eta_\ast \approx 50$), shows a distinctly lower level of such random variation in H$\alpha$ equivalent width
\citep{How2007}.
This suggests that the general number and size of structures may depend on, e.g., the size or rigidity of the magnetosphere.
This again requires further investigation through a systematic parameter study.

Finally, while the overall agreement between 3D model predictions and observed H$\alpha$ variations is quite good,  this is achieved with a modest adjustment in the assumed mass loss rate and/or wind clumping factor. Future work should examine the role of the line-deshadowing instability in inducing clumping in the wind acceleration region, and how well the numerical grid resolves this and other small-scale flow structures arising from wind compression in closed magnetic loops. 
In addition, future 3D models should consider both dipolar and non-dipolar field topologies  that are not aligned to the rotation axis. 
 In summary, while the results here have provided an encouraging first step, there remains much work to develop realistic 3D models of the magnetospheres and channeled wind outflows from magnetic massive stars.

\section*{Acknowledgments}
This work was carried out with partial support by NASA ATP Grants  NNX11AC40G  
and NNX08AT36H S04.

\bibliography{Ha_t1oc}

\end{document}